\documentclass[aps,reprint,a4paper,showpacs,amsmath,amssymb,superscriptaddress,floatfix]{revtex4-1}
\usepackage{amssymb}
\usepackage{latexsym}
\usepackage[dvips]{graphicx}
\usepackage{epsfig}

\usepackage{dcolumn}
\usepackage{amsmath}
\usepackage{amsfonts}
\usepackage{bm}
\usepackage{color}

\usepackage{hyperref}
\begin{document}
\title{Quantum Engineering of Superdark Excited States in Arrays of Atoms}
\author{A. A. Makarov$^{1,2,3}$, and V. I. Yudson$^{3,1}$}
 \affiliation{
$^{1}$Institute of Spectroscopy, Russian Academy of Sciences,
5 Fizicheskaya St., Troitsk, Moscow 108840, Russia \\
  $^{2}$Moscow Institute of Physics and Technology, Institutskiy
 pereulok 9,
Dolgoprudny, Moscow Region 141700, Russia \\
$^3$ National Research
University
Higher School of Economics, 20 Myasnitskaya St., Moscow 101000, Russia}
\date{\today}
\begin{abstract}
We suggest a regular method of achieving an extremely long lifetime of a
collective singly excited state in a generic small-size ensemble of $N$ identical
atoms. The decay rate $\Gamma_N$ of such a `superdark' state
can be as small as $\Gamma_N \propto \Gamma (r/\lambda)^{2(N-1)}$
($\Gamma$ is the radiative decay rate of an individual atom, $r$
and $\lambda$ are the system size and the wavelength of the
radiation, respectively), i.e., considerably smaller than in any of the
systems suggested up to now.
The method is based on a special fine tuning of the atomic Hamiltonian: namely,
on a proper position-dependent adjustment of atomic transition frequencies.
So chosen set of the control parameters is sufficient to ensure
the minimum of the spontaneous decay rate of the engineered state
in a generic ensemble of atoms (`qubits').
\end{abstract}

\maketitle
The phenomenon of subradiance (and superradiance as
well) is one of the central points when discussing
emission properties of
systems of identical
closely spaced atoms (qubits). There is growing
interest in sub-wavelength atomic ensembles with
an enhanced radiative lifetime of collective
excitations. Slowly decaying (dark) states can be implemented
for the storage of information in quantum memory devices.

Several configurations of one- and two-dimensional (1D and 2D)
atomic arrays have been proposed (see, e.g.
\cite{ZOUBI,SUTHERLAND,ASENJO,FACCHINETTI,PLANKENSTEINER,ZHANG,KORNOVAN}
that show substantial decrease of the decay rate as compared to that
of an individual atom.
For two atoms, a few  schemes  for
controlling  subradiant  states  in 1D \cite{ML,RY,BARANGER},
and in 3D \cite{ZADKOV,DAS,MY} were suggested.
An  interesting  (although  rather  complicated)  scheme for constructing a
singly excited subradiant state in a one-dimensional
array of many atoms  was  considered  in
Refs.~\cite{SCULLY1,SCULLY2}. As for the experiment,
some evidence in favour  of  a  change  in  the  spontaneous  decay  rate  was
obtained,  for  example,  in Ref.~\cite{PAVOLINI} for  an
 ensemble  of  many
 atoms and in
Ref.~\cite{BREWER} for a system  of  two ions  in  a  trap.
Finally, subradiance from a cloud of cold atoms was reliably
observed  \cite{GUERIN,GUERIN2}.

The suppression of the radiative decay is caused by destructive
interference of emission amplitudes of different members of the atomic
ensemble. This effect is most pronounced in the seminal Dicke model
\cite{DICKE} of a compact ensemble of $N$ identical two-level atoms
without non-retarded dipole-dipole interaction:
one of the singly excited collective
states is super-radiative, while the other N-1 singly
excited states are non-radiative at all.

The presence of the  dipole-dipole interaction
leads to formation of the eigenstates (excitons) of
the atomic ensemble which may reveal the properties
of super- and subradiation. In small-size ensembles
the resonant dipole-dipole excitation transfer is the
strongest effect, while the spontaneous radiative decay
occurs due to a weaker interaction with the transverse
quantum electromagnetic field.
Both the exciton states and
their radiative decay rates are determined by the geometry of the atomic
array. The search for an array geometry with a minimal decay rate
seems to be the matter of art, intelligence, and luck.

    In this Letter we suggest a regular
method to achieve an extremely long lifetime of a collective state in a
generic small-size array of $N$ atoms. The decay rate $\Gamma_N$ of such
a `superdark' state can be as small as
$\Gamma_N \propto \Gamma (r/\lambda)^{2(N-1)}$
($\Gamma$ is the radiative decay rate of an individual atom, $r$ and $\lambda$
are, correspondingly, the system size and the resonant wavelength of the atomic transition).
The method is based on a special fine tuning of the exciton Hamiltonian.
We show that it is sufficient to adjust the frequencies of individual
atomic transitions.
We begin from an analysis of a generic atomic
ensemble and after that we concentrate on the case of a finite
regular  chain of atoms.

We consider an ensemble of $N$ identical atoms located at spatial points $\mathbf{R}_j$ ($j=1, 2, ..., N$).
Each of singly excited collective atomic states $|\mathbf{C}\rangle$ can be
represented as a superposition $|\mathbf{C}\rangle = \sum^N_{j=1}C_j|e; j\rangle$ of
basis states $|e; j\rangle$ where the $j$-th atom is in its excited state $|e\rangle$ while all
the other atoms are in the ground state $|g\rangle$.
These exciton states are eigenstates of the atomic ensemble Hamiltonian
$\widehat{H}^{({\mathrm{at}})}$:
$\widehat{H}^{({\mathrm{at}})}|\mathbf{C}\rangle = E |\mathbf{C}\rangle$, or in the matrix form
(we put $\hbar = 1)$
\begin{equation}\label{H}
\omega_jC_j + \sideset{}{'}\sum^N_{j'=1} U(\mathbf{R}_j, \mathbf{R}_{j'})C_{j'} = E C_j
\, .
\end{equation}
Here the prime at the sum symbol means the
exclusion of the term $j' = j$;
$\omega_j$ is the transition frequency of the $j$-th atom;
$U(\mathbf{R}_j, \mathbf{R}_{j'})$ is the matrix element
of the dipole-dipole interaction between the states $|e;j\rangle$
and $|e;j'\rangle$; its particular form will be specified below.
For the introductory illustration of the suggested approach we assume that
the excited atomic state is non-degenerate. Then we will describe a more general
situation. The radiative decay rate of a given exciton state is
determined by the Fermi Golden Rule:
\begin{multline}
\Gamma_N \propto \Gamma \sum_{\nu=1,2}\int d\mathrm{o}_{\mathbf{k}} \left| \sum^N_{j=1}
(\mathbf{d} \cdot \mathbf{e}^{(\nu)}_{\mathbf{k}})e^{i\mathbf{k}\mathbf{R}_j} C_j \right|^2
\\
= \Gamma \sum^N_{j,j'=1}C^*_j W_{j,j'}C_{j'} \, , \label{Gamma-W}
\end{multline}
where $\mathbf{d}= \langle e|\hat{\mathbf{d}}|g\rangle$  is the matrix element of the dipole moment operator and
$\mathbf{e}^{(\nu)}_{\mathbf{k}}$ is the unit polarization vector of
a photon of the wave vector $\mathbf{k}$. The integration in Eq.~(\ref{Gamma-W})
is performed over the solid angle in the momentum space
while $k = |\mathbf{k}|=2\pi/\lambda $ is fixed by the energy of the state $|\mathbf{C}\rangle$.
The first sum runs over the polarizations.
The quadratic form in
Eq.~(\ref{Gamma-W}) is positively defined. The matrix
$W_{j,j'}$ is obviously Hermitian. Moreover, it is real due to the
symmetry of the integrand in Eq.~(\ref{Gamma-W}) with respect to
the change $\mathbf{k} \rightarrow -\mathbf{k}$.
Therefore, for a given arrangement of atoms (i.e., for a given matrix
$W_{j,j'}$), there exists a real unit vector $\{C_j\}$
($j=1, ..., N; \,\,\,  \sum_j C^2_j =1$) that provides the minimal value
of the quadratic form and, therefore, realizes the minimal decay rate.
If the corresponding state $|\mathbf{C}\rangle = \sum^N_{j=1}C_j|j\rangle$
were an eigenstate of the atomic Hamiltonian $\widehat{H}^{({\mathrm{at}})}$,
the problem of finding the state with the
slowest decay rate would be solved.
However, in general,
the two matrices, $W_{j,j'}$ (\ref{Gamma-W}) and $H^{({\mathrm{at}})}_{j,j'}$ (\ref{H})
do not have a common eigenvector. Our current task is to perform a fine tuning
of the atomic Hamiltonian in order to make the state $|\mathbf{C}\rangle$
its eigenstate obeying the system of equations (\ref{H}).
In these equations the matrix $U(\mathbf{R}_j, \mathbf{R}_{j'})$ of
the dipole-dipole interactions is fixed by the positions of the atoms,
while the local frequencies $\omega_j$ can be
controlled by external fields. It is naturally to count these shifts
(and the eigenenergy $E$ as well) from an ensemble averaged frequency
$\omega_0$, so that $\sum^{N}_{j=1}\omega_{j}=0$.
This condition together with $N$ equations (\ref{H}) with fixed $\{C_j\}$
uniquely determine $N+1$ real quantities: local frequency shifts $\omega_j$ and
the eigenenergy $E$, as
\begin{eqnarray}
&& E =  \frac{1}{N} \sideset{}{'}\sum^N_{j,j'=1}\frac{1}{C_j} U(\mathbf{R}_j, \mathbf{R}_{j'})C_{j'}
\, ,  \label{E}  \\
&& \omega_j = E - \frac{1}{C_j} \sideset{}{'}\sum^N_{j'=1} U(\mathbf{R}_j, \mathbf{R}_{j'})C_{j'}
\, . \label{omega-j}
\end{eqnarray}
This solves the problem of the Hamiltonian tuning to ensure
the existence of an eigenstate with the minimal decay rate for the given geometry
of the atomic ensemble.

The considered model can be generalized to the case where
the non-degenerate ground atomic state
has the angular momentum $J=0$ while the excited state is
degenerate having the angular momentum $J=1$ with
projections $m=0, \pm 1$ on the quantization axis $\hat{z}$.
Hereafter we shall
use an equivalent basis of atomic excited states
$|\alpha\rangle$ ($ \alpha = x, \, y, \, z $), connected
with the states $|m\rangle$ by $|m=0\rangle = |z\rangle$,
$|m=\pm 1\rangle = ( |x\rangle \pm i|y\rangle )/\sqrt{2}$.
Matrix elements of the Cartesian components $\hat{d}_{\beta}$
($\beta = x, y, z$) of the dipole moment operator in this
representation are
$\langle \alpha |\hat{d}_{\beta} |g\rangle = d \delta_{\alpha,\beta}$.
The operator of the dipole-dipole interaction between
the atoms is
\begin{equation}\label{d-d}
\widehat{H}_d = \sideset{}{'}\sum^N_{j,j'=1}\frac{(\hat{\mathbf{d}}_j \cdot \hat{\mathbf{d}}_{j'}) -
3 (\hat{\mathbf{d}}_j \cdot \mathbf{n}_{j,j'})(\hat{\mathbf{d}}_j'\cdot \mathbf{n}_{j,j'})}
{|\mathbf{R}_j - \mathbf{R}_{j'}|^3}
\, ,
\end{equation}
where $\hat{\mathbf{d}}_j$ is the dipole operator of the $j$-th atom;
the unit vector
$\mathbf{n}_{j,j'}= (\mathbf{R}_j - \mathbf{R}_{j'})/|\mathbf{R}_j - \mathbf{R}_{j'}|$.
In general, the operator (\ref{d-d}) mixes excited states with different projections $\alpha$.
Therefore, collective singly excited eigenstates of the atomic array
are described by a superposition
\begin{equation}\label{C}
|\mathbf{C}\rangle = \sum_{\alpha} \sum^N_{j=1}C^{(\alpha)}_j|\alpha;j\rangle \;,\;\; \; \;
\sum_{\alpha} \sum^N_{j=1}|C^{(\alpha)}_j|^2 = 1
\, .
\end{equation}
Here the sum runs over $\alpha=x, y, z$; the basis vector $|\alpha;j\rangle$ denotes
the state where the $j$-th atom is in its excited
state $|\alpha\rangle$ ($J=1$) while all the other atoms are in their ground states ($J=0$). Correspondingly,
the equations for the eigenstates and for the radiative decay rate look similar to Eqs.(\ref{H}) and
(\ref{Gamma-W}) but with the matrices
$U^{\alpha,\beta}(\mathbf{R}_j, \mathbf{R}_{j'})$ and $W^{\alpha,\beta}_{j,j'}$ having the additional indices.
Interaction of the $j$-th atom with the plane wave of wave vector $\mathbf{k}$ and polarization $\nu$
is governed by the operator $H_{int} \propto - (\hat{\mathbf{d}}\cdot \mathbf{e}^{(\nu)}_{\mathbf{k}})
e^{i\mathbf{k} \cdot \mathbf{R}_j}$. The spontaneous decay rate of the state (\ref{C}) is given by
\begin{multline}\label{Gamma-J}
\Gamma_N =\frac{3\Gamma}{8\pi}\int\limits_0^{2\pi}d\varphi
\int\limits_0^\pi \sin\theta\,d\theta \sum^2_{\nu =1}
\left|\sum\limits_{j=1}^N\sum_{\alpha}
e^{(\nu)}_{\mathbf{k}, \alpha}C^{(\alpha)}_j e^{i\mathbf{k} \cdot \mathbf{R}_j}\right|^2  \\
 \equiv \Gamma \sum_{\alpha,\beta}\sum^N_{j,j'=1}C^{(\alpha)*}_j W^{\alpha,\beta}_{j,j'}C^{(\beta)}_{j'}
\, ,
\end{multline}
where the wave vector is parameterized as $\mathbf{k}= k(\sin\theta\cos\varphi,\;\sin\theta\sin\varphi\;, \cos\theta)$.
The polarization vectors are
chosen as $\mathbf{e}^{(1)}_{\mathbf{k}} = (-\sin\varphi, \cos\varphi, 0)$
and  $\mathbf{e}^{(2)}_{\mathbf{k}} =
(\cos\theta\cos\varphi,\cos\theta\sin\varphi, -\sin\theta)$; their Cartesian
components obey the relation
$\sum^2_{\nu=1} e^{(\nu)}_{\mathbf{k}, \alpha}e^{(\nu)}_{\mathbf{k}, \beta} =
\delta_{\alpha,\beta} - k_{\alpha}k_{\beta}/k^2$.

Using Eq.~(\ref{Gamma-J}) one can calculate the decay rate of any singly excited
eigenstate of the atomic Hamiltonian.
The matrix $W^{\alpha,\beta}_{j,j'}$ is Hermitian, and real
due to the symmetry $\mathbf{k} \rightarrow - \mathbf{k}$. The
matrix $U^{\alpha,\beta}(\mathbf{R}_j, \mathbf{R}_{j'})$ corresponding to
the matrix elements of the Hamiltonian (\ref{d-d}) between the states
(\ref{C}) is obviously real, too. Thus, similarly to
the simplified model, we can find the real vector
$|\mathbf{C}\rangle $ that
realizes the minimum of the quadratic form (\ref{Gamma-J}). Then
we can adjust the local atomic transition frequencies making this
optimal vector to be an eigenvector of the atomic Hamiltonian.
In this way, the conceptual problem of the proper tuning of the atomic
Hamiltonian to ensure the existence of an eigenstate with
the minimal (for the given array geometry) decay rate is solved.

For atomic ensembles of a low symmetry (where exciton
states are not characterized by a single atomic polarization
$\alpha$) implementation of the described procedure may be
complicated because it would require independent tuning of all
three initially degenerate atomic transition frequencies
of each atom. The situation simplifies for geometries where the
dipole-dipole interaction (\ref{d-d}) does not mix excited
states of the atomic Hamiltonian with different polarizations $\alpha$.
This happens in configurations where the vectors connecting atom sites
are parallel or perpendicular to the direction of the transition dipole
moment of a chosen excitation polarization.
Such are, for instance, a plane ensemble and
a linear chain of atoms.
Now, we study a linear chain of atoms
located at the sites $\mathbf{R}_j = R_j\hat{z}$.
We will explicitly construct the `optimal' vector $|\mathbf{C}\rangle$
and the atomic Hamiltonian so that the decay rate of the found `superdark'
state will be enormously small
$\Gamma_N \propto \Gamma (r/\lambda)^{2(N-1)}$.
For the excitation polarized parallel
(perpendicular) to the chain directions, i.e., for
$|\alpha\rangle =|z\rangle$
($|\alpha\rangle =|x\rangle$ or $|\alpha\rangle =|y\rangle$), matrix
elements of the operator (\ref{d-d}) for the excitation transfer between the
sites $R_j$ and $R_{j'}$ are given by
\begin{equation}\label{U}
U^{\parallel}_{j,j'} = -2\frac{d^2}{R^3_{j,j'}} \;,\; \;
U^{\perp}_{j,j'} = \frac{d^2}{R^3_{j,j'}}
\, ,
\end{equation}
where $R_{j,j'}$ is the distance between the sites $j$ and $j'$ of the chain.
The expression (\ref{Gamma-J}) for the decay rate of the collective singly
excited state of the polarization $\alpha = z$
($\parallel$), or $\alpha = x, y$ ($\perp$), reduces to
\begin{equation}\label{Gamma-a}
\Gamma_N
=\frac{3\Gamma}{8\pi}\int
d\mathrm{o}_{\mathbf{k}}
f^{(\alpha)}(\theta, \varphi)
\left|\sum\limits_{j=1}^N C^{(\alpha)}_j e^{ikR_j\cos\theta}\right|^2
\, ,
\end{equation}
where $f^{(\alpha)}(\theta, \varphi) = 1 - k^2_{\alpha}/k^2$, i.e.,
$f^{(z)}(\theta, \varphi) = \sin^2\theta $ and $f^{(x)}(\theta, \varphi) =
1 - \sin^2\theta\cos^2\varphi$. The integration determines the corresponding matrix
$W^{\alpha, \alpha}_{j,j'}$ of the quadratic form (\ref{Gamma-J}):
\begin{eqnarray}
W^{z,z}_{j,j'} \equiv W^{(\parallel)}_{j,j'} &=&  3
\frac{\sin\xi - \xi\cos\xi}{\xi^3}
\; ; \label{W-par} \\
W^{x,x}_{j,j'} \equiv W^{(\perp)}_{j,j'} &=&  \frac{3}{2}
\frac{\xi\cos\xi + (\xi^2 -1) \sin\xi}{\xi^3}
\; , \label{W-perp}
\end{eqnarray}
where the notation $\xi = kR_{j,j'}$ is used for brevity.
Exact eigenvectors corresponding to the minimal eigenvalues
of these matrices can be determined numerically.
To make a simple analytical estimate of the minimal decay rate
for a short chain (of the length $r \ll \lambda$ ), it is natural
to expand exponents in Eq.~(\ref{Gamma-a})
in powers of the small quantities $ kR_j\cos\theta $:
\begin{equation} \label{Cn}
\left|\sum\limits_{j=1}^N C^{(\alpha)}_j e^{ikR_j\cos\theta}\right|
=
\left|\sum^{\infty}_{n=0}\frac{(ikr\cos\theta)^n}{n!}
\sum\limits_{j=1}^N \left(\frac{R_j}{r}\right)^n C^{(\alpha)}_j \right|
\, .
\end{equation}
To minimise this expression (therefore, $\Gamma_N$)
we require vanishing of $N-1$ first terms ($n=0, ..., N-2 $)
of the expansion in $(kr)^n$. This means imposing
$N-1$ (i.e., the maximal possible number) linear
constraints on the $N$-component unit vector
$|\mathbf{C}\rangle$:
\begin{equation} \label{CN}
\sum\limits_{j=1}^N \left(\frac{R_j}{r}\right)^n C^{(\alpha)}_j = 0 \;;\;\;\; n= 0, 1, \ldots, N-2
\, .
\end{equation}
Under conditions (\ref{CN}) the expansion (\ref{Cn}) begins with terms
$\propto (kr)^{N-1}$ and we arrive at the announced estimate for
the minimal decay rate $\Gamma_N \propto (kr)^{2(N-1)}$. This minimal
value realized at the optimal vector determined by (\ref{CN}), is very
fragile: a slightest deviation of the state $|\mathbf{C}\rangle$ from
the optimal one would involve terms of the expansion (\ref{Cn}) with
lower powers of $kr$ (i.e., of a much larger value). This results in
a strong sharpness of the decay rate dependences on tuning parameters
(see below Fig.~1 and Fig.~2 for particular examples).

A formal solution to the system (\ref{CN}) can be expressed in terms of the
$N \times N$ Vandermonde matrix $\mathbf{V}$
($V_{ij}=(R_j/r)^i$ ; $i=0, \ldots, N-1; \, j=1, \ldots, N $):
$C^{(\alpha)}_j = c(\mathbf{V}^{-1})_{j,N}$, where $c$ is determined by the
normalization condition.

Explicit results can be presented for a chain of equally spaced atoms
$R_j = (j-1)a$; $j=1, \cdots, N$. In this case the exact (normalized)
solution to (\ref{CN}) is expressed in terms of the binomial coefficients:
\begin{equation} \label{C-answer}
C^{(\alpha)}_j = (-1)^{j-1} \frac{[(N-1)!]^2}{(j-1)!(N-j)!\sqrt{(2N-2)!}}
\, .
\end{equation}
Note that this solution is either symmetric (for odd $N$) or antisymmetric
(for even $N$) with respect to the replacement $j \rightarrow N-j+1$:
\begin{equation} \label{Symmetry}
C^{(\alpha)}_j = (-1)^{N-1}C^{(\alpha)}_{N-j+1}
\, .
\end{equation}
The decay rate (\ref{Gamma-a}) of this optimal state
$|\mathbf{C}\rangle = \sum^N_{j=1}C^{(\alpha)}_j|\alpha; j\rangle$ is
\begin{equation} \label{Gamma-z-answer}
\begin{array}{c}
 \Gamma^{\parallel}_N \\
 \Gamma^{\perp}_N
\end{array}
\left\} \; = \frac{3\Gamma}{4N^2-1} \frac{[(N-1)!]^2}{(2N-2)!}(ka)^{2(N-1)} \; \right\{ \begin{array}{c}
                                                                  1 \\
                                                                  N
                                                                \end{array}
                                                                \,
\end{equation}
where the symbol $\parallel$ ($\perp$) denotes the excitation polarization
parallel (perpendicular) to the chain.
Now, to make the found optimal state $|\mathbf{C}\rangle $ an eigenstate of the
atomic Hamiltonian, we perform the fine tuning of the local atomic frequencies.
The adjusted atomic frequencies are given by Eq. (\ref{omega-j}), where the
matrix of the dipole-dipole interaction for the considered geometry takes
the form (see Eq. (\ref{U}):
$U^{\parallel}(\mathbf{R}_j, \mathbf{R}_{j'}) = -2d^2/(|j-j'|a)^3$ or
$U^{\perp}(\mathbf{R}_j, \mathbf{R}_{j'}) = d^2/(|j-j'|a)^3$.
The eigenenergy $E$ of the constructed state $|\mathbf{C}\rangle$ is given
by Eq.~(\ref{E}). As a consequence of Eq.~(\ref{Symmetry}),
the shifted atomic frequencies possess the symmetry  $\omega_j = \omega_{N-j+1}$.
The both decay rates (\ref{Gamma-z-answer}) are very small when the system size
is small as compared to the resonance wavelength, $(N-1)a \ll \lambda$.
The same condition is sufficient to treat the decay rate (\ref{Gamma-J})
of the state (\ref{C}) independently of other eigenstates of the atomic ensemble.

What is exciting---the property of `superdarkness' is {\textit{extremely sensitive}} to the shift(s) of transition frequencies \cite{QE}. This is illustrated by the results of numerical calculations for the chain of three and four equally spaced atoms. For three  atoms, the only adjusted
parameter is the shift of the transition frequency of the middle atom with respect to
those of the edge atoms: $\Omega = \omega_2-\omega_1$.
To characterize the subradiance dependence
on this shift we
introduce the quantity
${\widetilde{\Gamma}}(\Omega)$ that is the slowest decay rate
of those for all eigenstates.
The dependences of so defined ${\widetilde{\Gamma}}$
on $\Omega/U$ are shown in Fig.~1 for two values of
the parameter $(ka)^2$; here $U$ is the dipole-dipole interaction between the neighboring atoms:
$U^{\perp}=d^2/a^3$ and $U^{\parallel}=-2d^2/a^3$ [see Eq.~(\ref{U})].
Notice that the sign of the optimal shift is negative for the perpendicular
polarization (${\widetilde{\Gamma}}^{\perp}$) and positive for the parallel polarization
(${\widetilde{\Gamma}}^{\parallel}$).
One can see very sharp minima for all curves, their abscissas being  very close to the `optimum asymptotic value' $\widetilde{\Omega}/U$ defined by Eq.(\ref{omega-j}) with the eigenvector (\ref{C-answer}). For $N=3$, $\widetilde{\Omega}/U=-7/8$.
Meanwhile, the optimum is actually reached at $\Omega/U$ that is a little shifted
as can be seen in Fig.~1.
At, e.g., $ka=0.1$, the mismatch is $\Delta\Omega/U \approx 3.0\!\cdot\!\!10^{-3}$ for
and, in a good approximation, scales $\propto(ka)^2$.
At the same time, the corresponding decay rates lie (due to the {\textit{extreme sharpness}}) a few
 lower than those defined by asymptotic Eq.~(\ref{Gamma-z-answer}). E.g., the fall is $\approx2.64$  times for ${\widetilde{\Gamma}}_{N=3}^{\perp}$ at $ka=0.1$.
\begin{figure}[t]{\label{three}}
\includegraphics[scale=0.5]{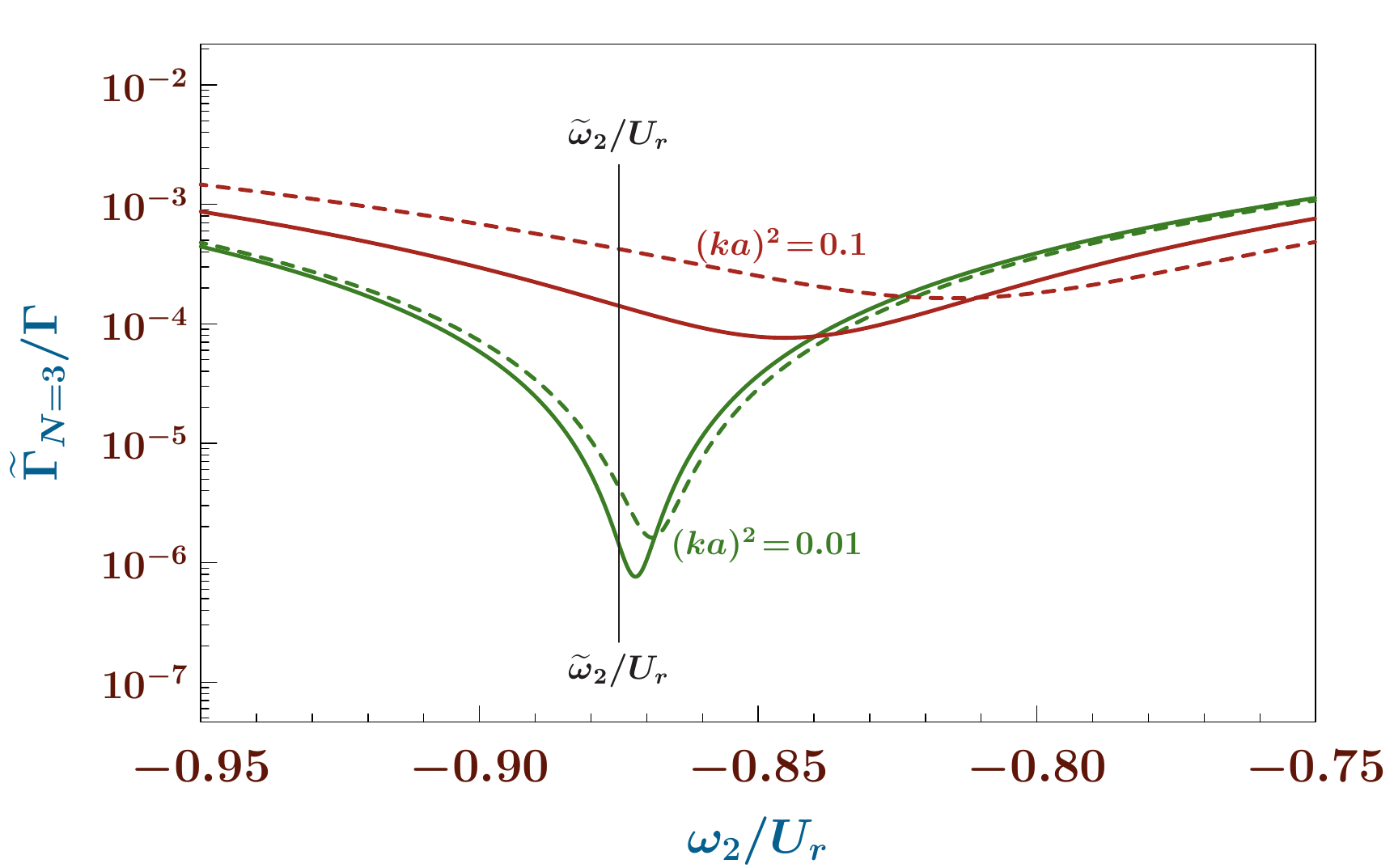}  
\caption{{\footnotesize{ (color online) Three atoms in chain configuration:
The slowest decay rates depending on the shift $\Omega/U$ of the transition frequency
for the middle atom with respect to the edge atoms (in the units of $U$, the dipole-dipole
interaction of the neighboring atoms, see details in the text).
The solid curves are for the polarization along the chain, and the
dashed curves are for the perpendicular one.}}}
\end{figure}

Similar results for four atoms are given in Fig.~2.
There is also only one adjusted parameter, i.e. the transition frequency shift $\Omega$
for the two middle atoms with respect to the edge atoms.
\begin{figure}[t]{\label{four}}
\includegraphics[scale=0.5]{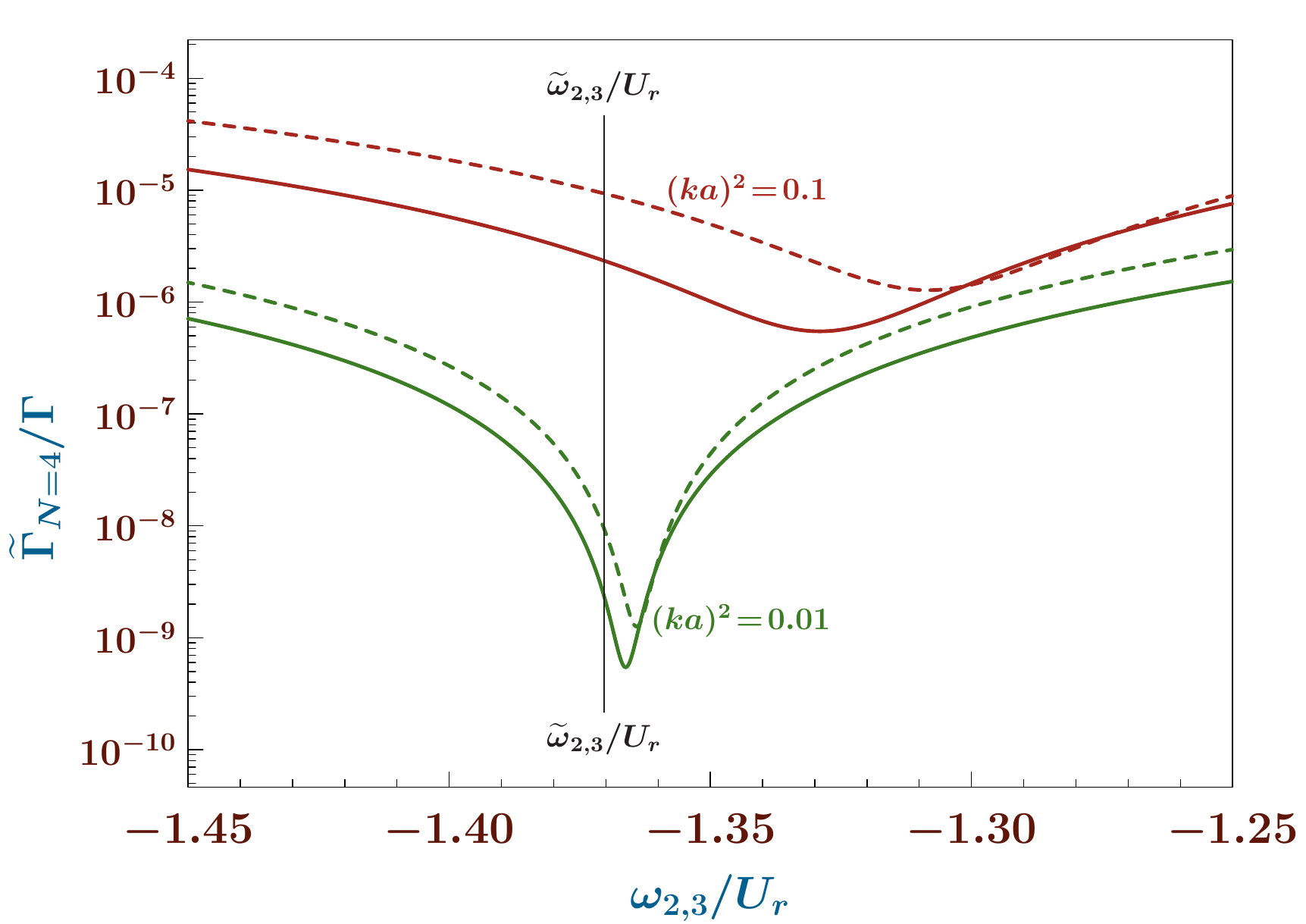} 
\caption{{\footnotesize{ (color online) Four atoms in chain configuration:
Notation is the same as in Fig. 1. A difference is that $\Omega/U = -37/27$.}}}
\end{figure}
In the case of $N\geqslant 5$ atoms the $[(N-1)/2]$ parameters must be adjusted. We don't draw the corresponding multidimensional pictures---only note that the minimum near the
set of optimum values defined by Eq. (\ref{omega-j}) with the eigenvector (\ref{C-answer}) is even sharper than in the just considered cases $N=3$ and $N=4$ (Figs. 1 and 2, respectively).

How strong could be the suppression of the decay rate ${\widetilde{\Gamma}}$ is shown in
Table 1,
where the values of ${\widetilde{\Gamma}}_N$ for the optimally adjusted atomic
frequencies are compared with those in the case where are no shifts, i.e. the transition frequencies of all atoms are equal.
\begin{table}
\caption{{\footnotesize{Comparison of two decay rates: The first one corresponds to
its minimum due to the suitable frequency shift $\Omega$  for $N=3$  and $N=4$; The second
one is the decay rate of the subradiative state in the case where are no shifts, i.e. the transition frequencies of all atoms are equal.}}}
\begin{tabular}{|c|c|c|c|c|}
 \hline
$N$ & ${\widetilde{\Gamma}}$ & $(ka)^2$ & At minimum\footnote{See Figs. 3 and 4.} & Without shift\footnote{All $\omega_j=0$.} \\ \hline
&& $0.01$ & $7.62\!\cdot\!\!10^{-7}$ & $0.0040$\\
{{$3$}} & ${\widetilde{\Gamma}}^\parallel/\Gamma$ & 0.10 & $7.64\!\cdot\!\!10^{-5}$ & $\!\!\!0.039$ \\
&& $1.00$ & $7.73\!\cdot\!\!10^{-3}$ & $\!\!\!\!\!\!\,0.036$ \\ \hline
&& $0.01$ & $1.62\!\cdot\!\!10^{-6}$ & $0.0079$ \\
{{$3$}} & ${\widetilde{\Gamma}}^\perp/\Gamma$ & 0.10 & $1.64\!\cdot\!\!10^{-4}$ & $\!\!\!\!\!\!\,0.056$ \\
&& $1.00$ & $1.79\!\cdot\!\!10^{-2}$ & $\!\!\!\!\!\!\,0.025$ \\ \hline
&& $0.01$ & $\;\,5.45\!\cdot\!\!10^{-10}$ & $4.4\!\cdot\!\!10^{-4}$ \\
{{$4$}} & ${\widetilde{\Gamma}}^\parallel/\Gamma$ & 0.10 & $5.48\!\cdot\!\!10^{-7}$ & $4.1\!\cdot\!\!10^{-3}$ \\ && $1.00$ & $5.78\!\cdot\!\!10^{-4}$ & $2.2\!\cdot\!\!10^{-2}$ \\ \hline
&& $0.01$ & $1.26\!\cdot\!\!10^{-9}$ & $8.8\!\cdot\!\!10^{-4}$ \\
{{$4$}} & ${\widetilde{\Gamma}}^\perp/\Gamma$ & 0.10 & $1.28\!\cdot\!\!10^{-6}$ & $8.0\!\cdot\!\!10^{-3}$ \\ && $1.00$ & $1.46\!\cdot\!\!10^{-3}$ & $2.7\!\cdot\!\!10^{-2}$ \\ \hline
  \end{tabular}
 \end{table}
The same several orders of magnitude are in play
comparing with the results for earlier treated subradiative systems
including a very recently suggested ensemble in the form of a regular
polygon \cite{CARDONER2} where $\Gamma_N\propto\Gamma(r/\lambda)^{2[N/2]}$.
The two exceptions are: (i) two separated by the distance $n\lambda/2$  atoms in 1D \cite{ML,RY};
(ii) extension of  this idea to 3D using two similarly separated two-dimensional arrays of atoms \cite{GUIMOND}. However, the chain configuration is much more commonly considered due to its implementation using the optical lattices, even subwavelength ones \cite{DALIBARD,WANG}. Our examples above relate
just to such a scheme at $ka\leqslant1 $ (for matching with $ka>1$ see Ref.~\cite{SHEREMET}).
Certainly,  usage of different tweezers for different atoms can be assumed. In addition we may suggest, in view of the both $N=3$ and $N=4$ cases, a speculative scheme of laser control using the standing wave as shown in Fig.~3.
\begin{figure}[t]
\includegraphics[scale=0.5] {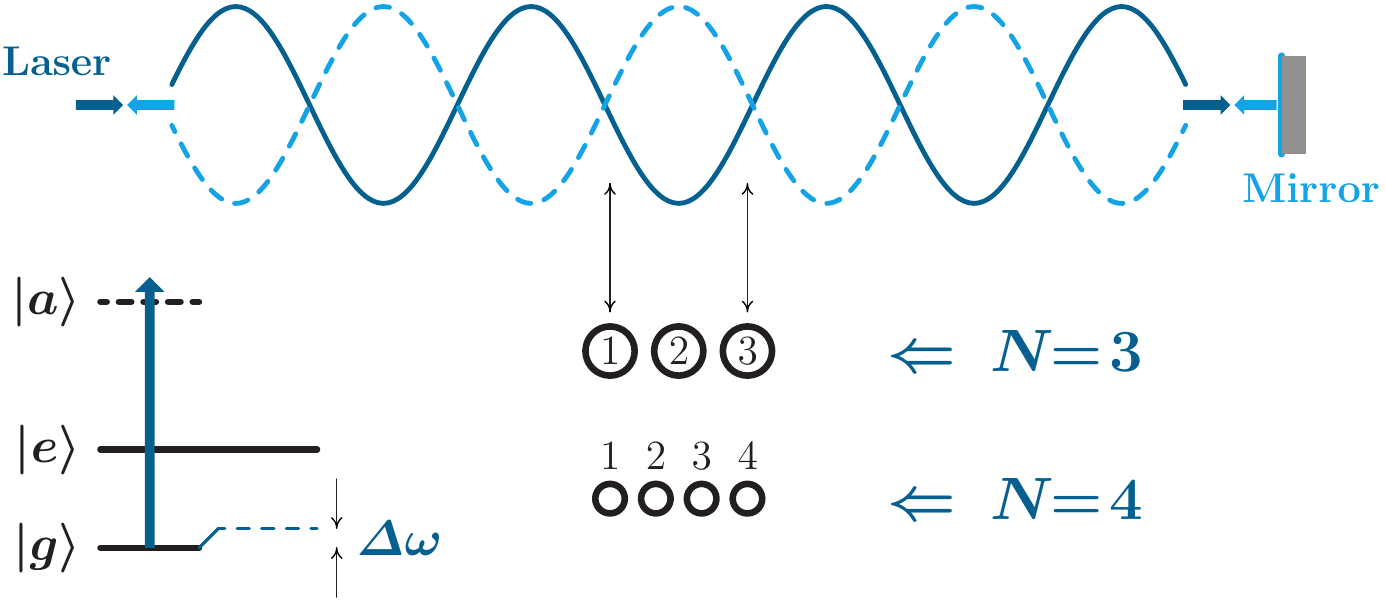} 
\caption{{\footnotesize{ (color online) Scheme for inducing an energy shift of the ground state in the middle atom(s). The nodes are at the positions of the edge atoms. The  laser  frequency is detuned relative to the frequency of the transition from the
ground state $|g\rangle$
to some high-lying excited state $|a\rangle$  of the atom(s). As shown in the diagram, this shifts the frequency of the transition $|e\rangle\rightarrow|g\rangle$ to the red, i.e. suitably for the perpendicular polarization. For the case of the parallel polarization the sign of the detuning should be reversed.
}}}
\end{figure}

To conclude,
in this Letter the recipe is presented how to achieve a huge gain in subradiance
of atomic ensembles. This gain is reached in a regular way by adjustment of
atomic transition frequencies $\omega_j$.
The demonstrated extreme sensitivity of the slow spontaneous decay rate and
the accompanying narrow radiative width to external fields
may be useful for precision measurements and diagnostics in addition to its undoubted
fundamental significance.
\vspace{0.2cm}

V.I.Yu. acknowledges partial support
from the Basic research program of HSE.


\begin{thebibliography}{99}
\bibitem{SHEREMET}In Ref.~\cite{KORNOVAN}
an extraordinary small $\Gamma_{N=10}^\perp\!\approx\!10^{-5}\Gamma$ was numerically discovered  at
$ka\!\approx\!0.23\!\cdot\!2\pi\!\approx\!1.44$, i.e. $ka>1$. Our calculation using the eigenvector (\ref{C-answer}) gives a greater value of
$\Gamma_{N=10}^\perp\!=\!2.8\!\cdot\!10^{-4}\Gamma$;  however, at $ka=1$ it turns out to be
$7.8\!\cdot\!10^{-7}\Gamma$, i.e. much less than that of Ref.~\cite{KORNOVAN}.
\end{thebibliography}
\end{document}